\documentclass[12pt,preprint]{aastex}
\shorttitle{A New Globular Cluster}
\shortauthors{Strader \& Kobulnicky}
\def\etal{{\it et al.}}
\usepackage{subfigure}

\begin{document}

\title{A Probable New Globular Cluster in the Galactic Disk}

\author{Jay Strader\altaffilmark{1,2}, Henry A.~Kobulnicky\altaffilmark{3}}
\email{jstrader@cfa.harvard.edu, chipk@uwyo.edu}

\altaffiltext{1}{Harvard-Smithsonian Center for Astrophysics, Cambridge, MA 02138}
\altaffiltext{2}{Hubble Fellow}
\altaffiltext{3}{Department of Physics \& Astronomy, University of Wyoming, 1000 E. University, Laramie, WY 82071}

\begin{abstract}

We report the discovery of a probable new globular cluster in the disk of the Milky Way. Visible in 2MASS 
and the GLIMPSE survey, it has an estimated foreground extinction of $A_V \sim 24$ mag. The absolute 
magnitude of the cluster and the luminosity function of the red giant branch are most consistent with that 
of an old globular cluster with a mass of a few $\times 10^{5}  M_{\odot}$ at a distance of 4--8 kpc.

\end{abstract}


\section{Introduction}

Harris (2001) estimated that there were $\sim 20$ unknown Galactic globular clusters hidden behind substantial 
foreground extinction in the disk or behind the bulge. Subsequent near-IR surveys of the disk have borne out the 
prediction of missing clusters, two of which have been discovered in 2MASS by Hurt \etal~(2000). Another cluster, 
GLIMPSE-C01, was found by Kobulnicky \etal~(2005) using the Spitzer/IRAC GLIMPSE survey (Benjamin \etal~2003) of 
the Galactic plane. The importance of the Galactic globular cluster system in understanding the formation, 
evolution, and destruction of globular clusters motivates continuing efforts to finish the census of clusters.

In this paper we report the discovery of a probable globular cluster at Galactic coordinates $l=14.13$, 
$b=-0.64$ (J2000 coordinates: $18^{\textrm{h}} 18^{\textrm{m}} 30^{\textrm{s}}$ --$16^{\circ} 58\arcmin 
36\arcsec$). This object was 
identified by Mercer \etal~(2005) in a search for star clusters in GLIMPSE, although it was not 
suggested to be a globular cluster. As this cluster is \#3 in their catalog, we refer to the object as 
Mercer 3. This is relatively consistent with the rather confused naming conventions for Galactic 
globular clusters.

\section{Imaging}

Figure 1 and 2 show 2MASS $JHK$ images of a $2.4\arcmin \times 2.4\arcmin$ area around Mercer 3, as well as 
a red DSS image on the same scale. The cluster is clearly visible in $H$ and $K$ but disappears in $J$, 
indicative of the high foreground reddening. There is no evidence of the cluster in the DSS image. In Figure 
3 a composite color image of IRAC Band 2 ($4.5 \micron$), Band 4 ($8 \micron$), and MIPS $24 \micron$ is 
shown. The morphology of the cluster is similar in the IRAC bands to that in $H$ and $K$. Patchy, diffuse 
emission is visible across the frame. However, there is no evidence of a bubble or shell that could suggest 
a young cluster.

An elongated, opaque cloud dominates the composite image, located only 3\arcmin~in projection from the 
cluster. Smaller dark clouds are located across the frame. We hypothesize that this large cloud complex is 
located in the foreground of the cluster and is the primary cause of the large extinction we derive later in 
the paper.

Using the 2MASS images, we performed integrated aperture photometry within a radius of 75\arcsec centered on 
the cluster, using a concentric sky aperture between 75 and 90\arcsec. We do not claim that 75\arcsec is 
certain to be the edge of the cluster in a meaningful sense, but the extinction becomes noticeably variable 
at larger radii and so a larger aperture cannot be used. Thus light in the outermost parts of the clusters 
will be lost. A competing effect is that our sky aperture may have larger extinction than the inner parts of 
the cluster, leading to an undersubtraction of the background and so an overestimate of the cluster 
luminosity. It is difficult to assess the relative importance of these two effects and we caution the reader 
that our total magnitudes are likely to be uncertain at least at the 0.2--0.3 mag level.

We obtain total integrated magnitudes of $H=7.3$ and $K=6.1$. The half-light radius of the cluster in $K$ is 
$\sim 39$\arcsec, but this is a lower limit due to the uncertainty in the amount of light at large radii and 
the extinction gradient in the image. For the likely distance range derived in \S 4, 4--8 kpc, this 
corresponds to a half-light radius between 0.8 and 1.5 pc. The low end of the range is smaller than nearly 
all Galactic GCs, while a value of 1.5 pc is smaller than typical but not unusual. It is noteworthy that 
GLIMPSE-C01 and another recent Galactic plane discovery FSR 1767 (Bonnato \etal~2007) also have very small 
half-light radii of $\sim 0.6-0.7$ pc (Kobulnicky \etal~2005; Bonatto \& Bica 2008). It is unclear at 
present whether these small radii are accurate or are an artifact of the high extinction and the resultant 
difficulty in obtaining good surface brightness profiles.

\section{Stellar Photometry}

We use point source photometry taken from the Version 2.0 Data Release of GLIMPSE and matched with 2MASS 
sources. Within a radius of 30\arcsec from the cluster center, we select only those sources that are 
detected in all of $H$, $K$, and IRAC Bands 1 and 2. 25 stars fit these criteria. Figure 4 shows 
$K$ vs.~$H-K$ and $K$ vs.~$K-3.6$ color-magnitude diagrams as observed; no extinction corrections 
have been applied. 12 Gyr [$Z$/H] $= -2$ isochrones from Marigo \etal~(2008) are plotted, assuming a 
distance of 5 kpc and $E(B-V)=7.7$ (the values derived in \S 3.1 and \S 4).

Both panels show a broad column of stars extending across $\sim 3$ mag in $K$. The cutoff at $K \sim 
14$ is due to the photometric limit of 2MASS; the IRAC images go somewhat deeper. A reasonable 
assumption is that these stars are the brightest red giants in the cluster, and that the spread in 
colors is due to differential extinction.

\subsection{Extinction}

We can constrain the extinction toward the cluster by noting the remarkable fact that the IR colors of 
red giants vary little with age or metallicity, except at the tip of the red giant branch where there 
are few stars. The typical spread in colors is no more than $\sim 0.1$ mag in $H-K$ and 0.05 mag in 
$K-3.6$. Thus we can use a color-color plot to estimate the extinction with no knowledge of the 
metallicity or age of the cluster.

Figure 5 is an $K-3.6$ vs.~$H-K$ color-color plot. Overplotted are lines representing the reddened mean color of 
the upper red giant branch for a 12 Gyr old globular cluster; the isochrones used are Marigo \etal~(2008). Two 
extreme metallicities are plotted, $-2$ and 0. The reddenings range from $E(B-V)$ = 6 to 9, with crosses marking 
each magnitude of reddening. This figure shows that the unknown metallicity of the cluster has a minor effect on 
the red giant branch color compared to the spread in the points, suggesting that the differential extinction 
dominates the error. Since the lines do not pass directly through the center of points, one derives a different 
reddening from each of the colors. In $H-K$ the mean reddening appears to be $E(B-V) \sim 7.8$, compared to 
$E(B-V) \sim 7.5$ for $K-3.6$. The differential reddening is at least 1 mag in both colors, though this may be 
exaggerated by the contamination of our sample with field stars (the stars lying far from the central clump in 
Figure 5, for example, are unlikely to be cluster members). Stellar population models are better tested in the 
classic near-IR bands of $H$ and $K$ than in the newer Spitzer bands, so we tend to slightly favor the $H-K$ 
value. 
Thus we will adopt $E(B-V) \sim 7.7$ as our fiducial mean reddening, keeping in mind the presence of large 
differential reddening.

An $E(B-V)$ value of $\sim 7.7$ is extraordinarily high, corresponding to $A_K \sim 2.8$ and $A_V \sim 
24$. We can do a sanity check by noting that the cluster is not detected in the 2MASS $J$ image. The 
brightest red giants have $K \sim 11$, equivalent to an extincted $J \sim 16$. Below $J \sim 16$, 
especially in crowded regions, the completeness of 2MASS drops significantly, consistent with the 
absence of anything but a few stars at the position of the cluster in the $J$ image in Figure 1. 
However, if the reddening were as low as $E(B-V)= 5.5$ or 6, then the brighter red giants would be 
visible in $J$. We conclude that our derived reddening is consistent with the lack of the cluster in the 
2MASS $J$ image.

\subsection{The Source of the Extinction}

As discussed above, Mercer 3 is located several arcmin in projection from a large IR dark cloud. It is 
possible that this cloud is associated with the material responsible for the large reddening towards the 
star cluster.

In their discovery paper of a globular cluster in GLIMPSE, Kobulnicky \etal~(2005) used relatively high 
resolution CO data from the Galactic Ring Survey (Jackson \etal~2006) as a consistency check on the 
extinction toward their cluster. Unfortunately, Mercer 3 falls outside of the footprint of this survey, 
so we must fall back on older, lower-resolution data from the Massachusetts-Stony Brook Galactic Plane 
CO Survey (Clemens \etal~1986). The resolution of these data is $\sim$ 6\arcmin, too low to compare the 
morphology of the cloud in Figure 3 to the CO maps.

We downloaded a data cube from this survey covering the position of our cluster and extracted an 
integrated CO spectrum at the position of our cluster. The only significant feature is a strong peak at 
20 km/s. Integrating over the profile gives an intensity $I_{CO} = 93$ K km/s. This may be converted 
into an $H_2$ column density and optical extinction using the equations in Bohlin \etal~(1978; see also 
Kobulnicky \& 
Skillman 2008): $N_{H_2} = 3 \times10^{20} I_{CO}$ and

\begin{equation}
A_V = 3.1 \frac{2 N_{H_2}}{5.8 \times 10^{21}}
\end{equation}

Substitution yields $A_V \sim 30$ along this line of sight, generally consistent with the value derived 
from the color-color diagram. These equations assume that the CO is not optically thick and that there 
is no contribution of \ion{H}{1} to the extinction, and so represent a lower limit. On the other hand, 
some of the molecular gas may be behind the cluster; due to the low resolution of the data, the gas 
might also be associated with a different cloud that is in front of the cluster but not contributing to 
the foreground extinction.

If we assume that the CO cloud is predominately in the foreground, then we can use its velocity to 
constrain the near/far distance of the cluster. For $v = 20$ km/s and a Galactic $l = 14.1$, the 
near/far distances are 2.4 and 14.1 kpc. We can then, at the very least, take 2.4 kpc as a lower limit 
on the cluster distance. In the next section we will use the color-magnitude diagram of cluster stars to 
derive an upper limit on the distance.

\section{Luminosity Function, Age, and Distance}

An additional constraint on the distance and age of the cluster comes from the stellar luminosity function 
(LF). With only 25 stars in the complete sample, creating a useful $K$-band LF is impractical. However, 
the GLIMPSE data are deeper, so if we relax the restriction on matches with 2MASS, we can select a 
sample of stars with detections in IRAC bands 1 and 2. Within 30 \arcsec of the cluster center, there 
are 70 such stars. Figure 6 shows the 3.6$\mu$m LF plotted as a density estimate, using an Epanechnikov 
kernel and a bin width of 0.25 mag. The main features of the LF are: (i) a lack of stars brighter than 
$m_{3.6} = 9$, (ii) significant incompleteness below $m_{3.6} \sim 13$, and (iii) a gently upward 
sloping LF between these two limits. Overplotted are theoretical LFs from Marigo \etal~(2008) for solar 
metallicity and a range of ages from 1 Gyr to 12 Gyr (for old ages, the differences between metal-rich 
and metal-poor LFs in 3.6$\mu$m are small compared to the quality of our data and the effect of 
differential reddening). These have been scaled in distance and normalization to produce the best match 
for each age.

A generic feature of the LFs for ages younger than $\sim 2-3$ Gyr is a bump in the LF at the brightest 
magnitudes due to red supergiants. This bump is especially pronounced for ages of $\sim 1$ Gyr and for 
certain younger ages. No such feature is seen in the observed LF. Thus, independent of the cluster 
distance, the LF is inconsistent with Mercer 3 being a young star cluster. The LF is most consistent with 
that of a relatively old open or globular cluster.

Further constraints on the distance of the cluster come from the assumption of a particular age. For an 12 
Gyr solar metallicity stellar population, the maximum distance comes from identifying the brightest stars 
with the tip of the giant branch. This corresponds to an extincted distance modulus of $m-M \sim 16$, or 
$m-M_0 \sim 14.4$ using $A_{3.6} = 1.6$ (assuming $E(B-V) = 7.7$). Thus the maximal cluster distance is 
$\sim 7.6$ kpc. The shape and normalization of the LF appear to be somewhat better fit by a distance of 5.0 
kpc (this is the fit plotted in Figure 6), although the fit is far from perfect. Distances of 4 kpc or 
smaller are poor fits, as the theoretical LFs begin to rise steeply in a way unmatched by the data. This 
might partially be addressed by positing incompleteness at a brighter magnitude. Recall that the near 
distance limit from the CO data was 2.4 kpc. We conclude that a plausible distance range for an old cluster 
is 4--8 kpc, with a value closer to the middle of that range somewhat favored.

Assuming an age near the opposite extreme of the allowed range gives an upper limit on the distance. As 
shown in Figure 6, a 5 Gyr solar metallicity population appears to fit about as well as did the 12 Gyr LF. 
The implied distance is $\sim 12-13$ kpc (with a large error). What does this long distance imply for the 
mass of the cluster? Given the extinction and total $K$ magnitude discussed earlier, we derive $M_K \sim 
-12.1$ for a distance of 12 kpc. Using Maraston (2005) models with a Kroupa initial mass function, this is 
equivalent to a mass of $\sim 8 \times 10^5 M_{\odot}$. Mercer 3 would be one of the most massive star 
clusters in the Galaxy.

Alternatively, if we assume a distance of 5 kpc and an age of 12 Gyr, the implied mass is $\sim 2-3 \times 
10^5 M_{\odot}$ (depending on metallicity). This is close to the peak of the log-normal globular cluster 
luminosity function and would essentially peg Mercer 3 as a typical Milky Way globular cluster---keeping 
in mind that the total cluster luminosity is still quite uncertain.

As a conservative check on this mass estimate, we set aside the integrated $K$-band magnitude of the cluster for 
a moment and simply coadd the flux from all of the sources that lie along the red giant branch in Figure 4. This 
gives $K \sim 8.5$. We then make the almost absurd assumption that we have detected all of the red giants in the 
cluster (unlikely both because of incompleteness and because we are only considering sources within 30\arcsec of 
the center). Noting that standard stellar population models (e.g., Worthey 1994) predict that 50\% -- 60\% of the 
total $K$-band flux of an intermediate-age to old object will be from the red giant branch, we derive a total $K$ 
mag of $\sim 7.8$. For an old object at 5 kpc, this corresponds to a mass of $\sim 5 \times 10^{4} 
M_{\odot}$---less massive than a typical globular cluster, but not unusual, and still much more massive than 
nearly all open clusters.

Our conclusion from this line of argument is that Mercer 3 is most likely to be a typical old globular 
cluster, but we cannot rule out a less massive globular cluster or a more massive intermediate-age object.

\section{Discussion}

A secure identification of Mercer 3 as an old globular cluster will require moderately deep near-IR photometry. 
The predicted main sequence turnoff is at $K \sim 19$; if instead it is a massive intermediate-age cluster, the 
turnoff will be more than a magnitude fainter, and the shape of the subgiant branch will be significantly 
different. Due to the large and differential reddening, an accurate estimate of the cluster metallicity will 
probably require near-IR spectroscopy. This is easily accomplished as the brightest red giants have $K \sim 11$.

Mercer 3 is the second probable globular cluster discovered using Spitzer and 2MASS; a further two 
globular clusters were found using 2MASS alone. At the opposite end of parameter space, Koposov 
\etal~(2007) discovered two extraordinarily low-mass globular clusters in the Sloan Digital Sky Survey at 
heliocentric distances of $\sim 40-50$ kpc. The continuing pace of these discoveries suggests that the 
Galactic cluster census is far from complete.

\acknowledgments

J.~S.~was supported by NASA through a Hubble Fellowship, administered by STScI. We thank Dan Clemens, Laura 
Chomiuk, and Beth Willman for useful comments on the manuscript.

\newpage

\begin{figure}
\epsscale{0.85}
\plottwo{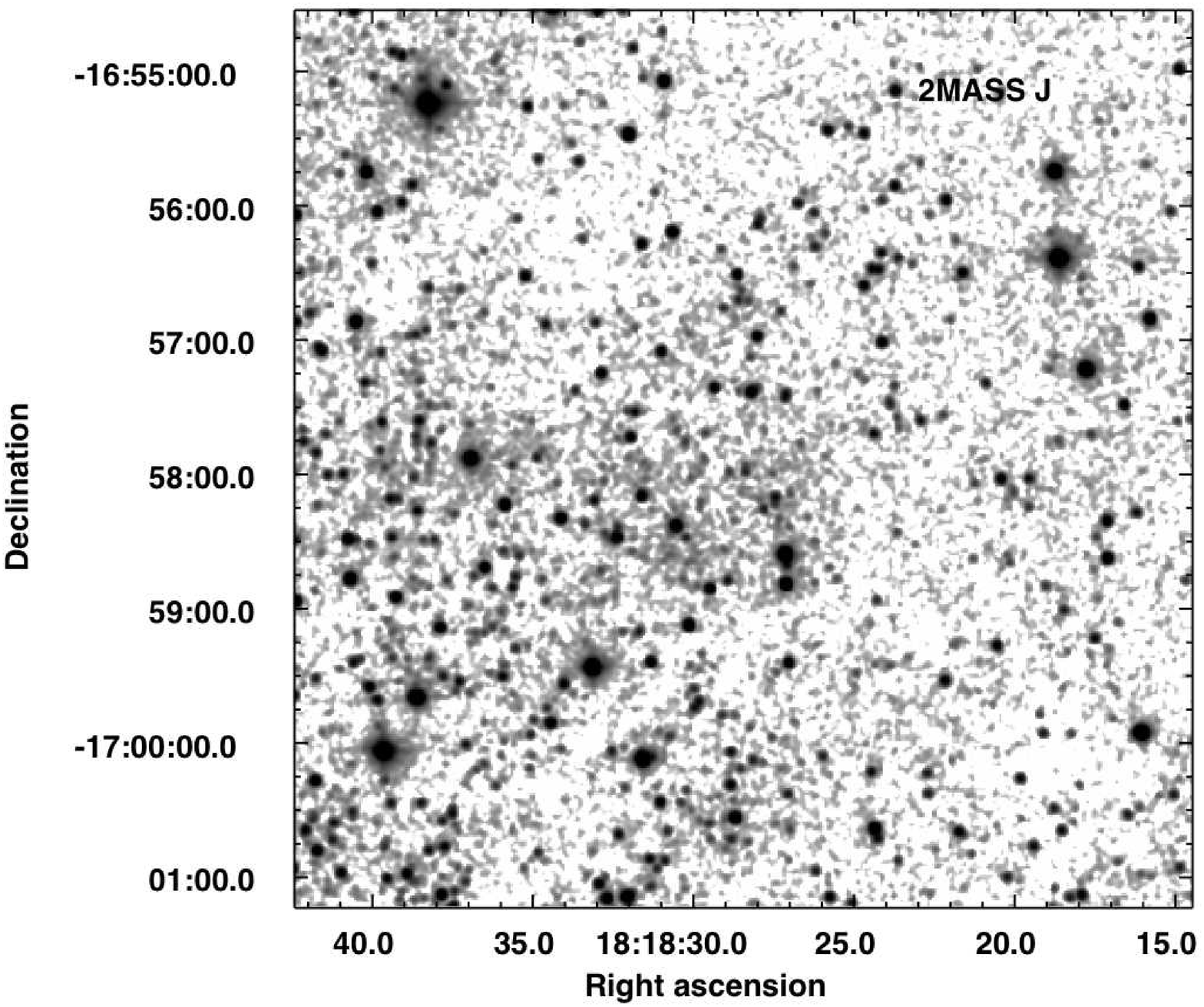}{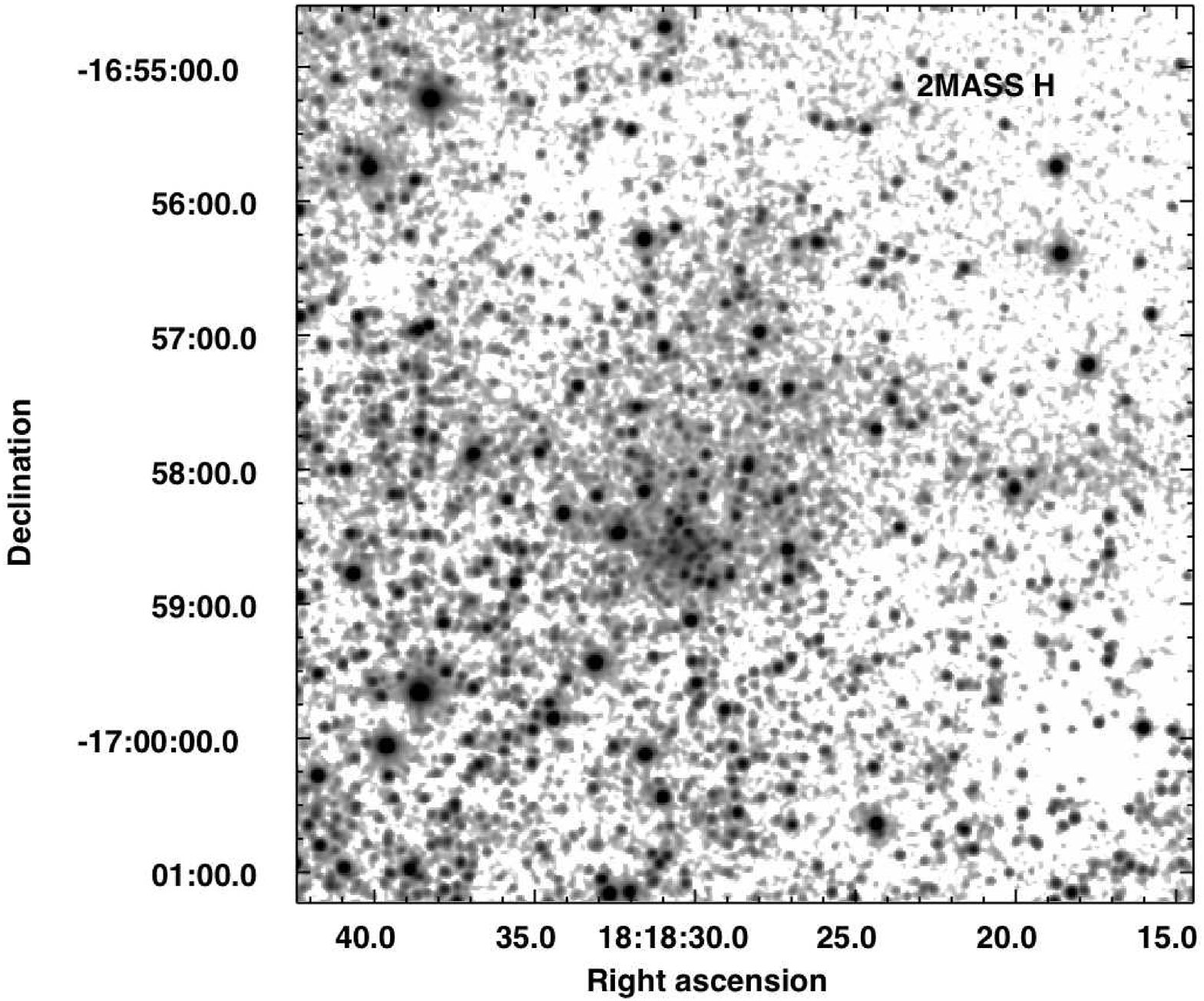}
\figcaption[f1.eps]{\label{fig:fig1}
Postage stamp images of the cluster in 2MASS $JH$. The cluster is
not visible in the $J$ image due to high extinction.}
\end{figure}

\begin{figure}
\epsscale{0.85}
\plottwo{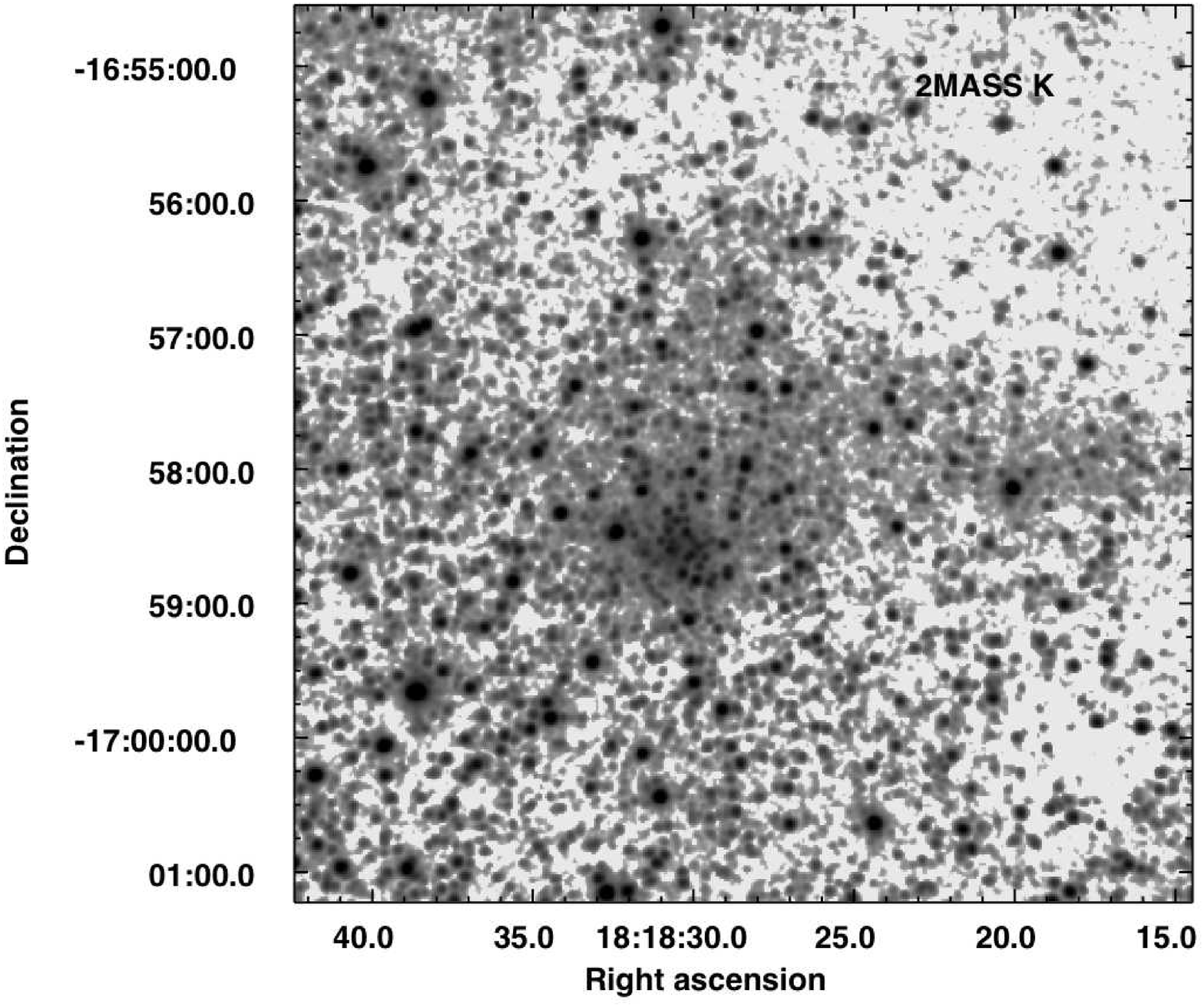}{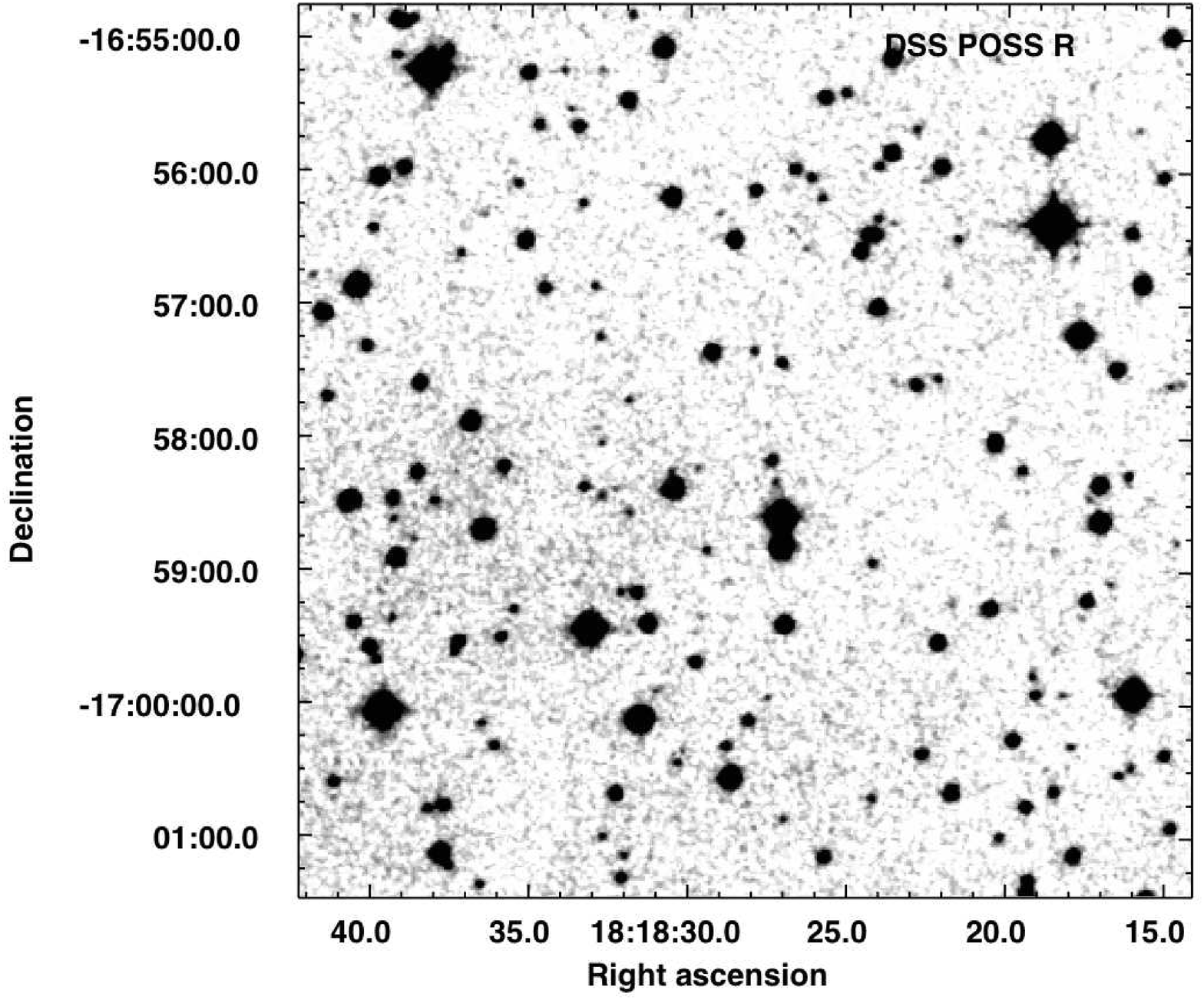}
\figcaption[f2.eps]{\label{fig:fig2}
Postage stamp images of the cluster in 2MASS $K$ and in red DSS. The cluster is
not visible in DSS due to high extinction.}
\end{figure}   

\begin{figure}
\epsscale{0.85}
\plotone{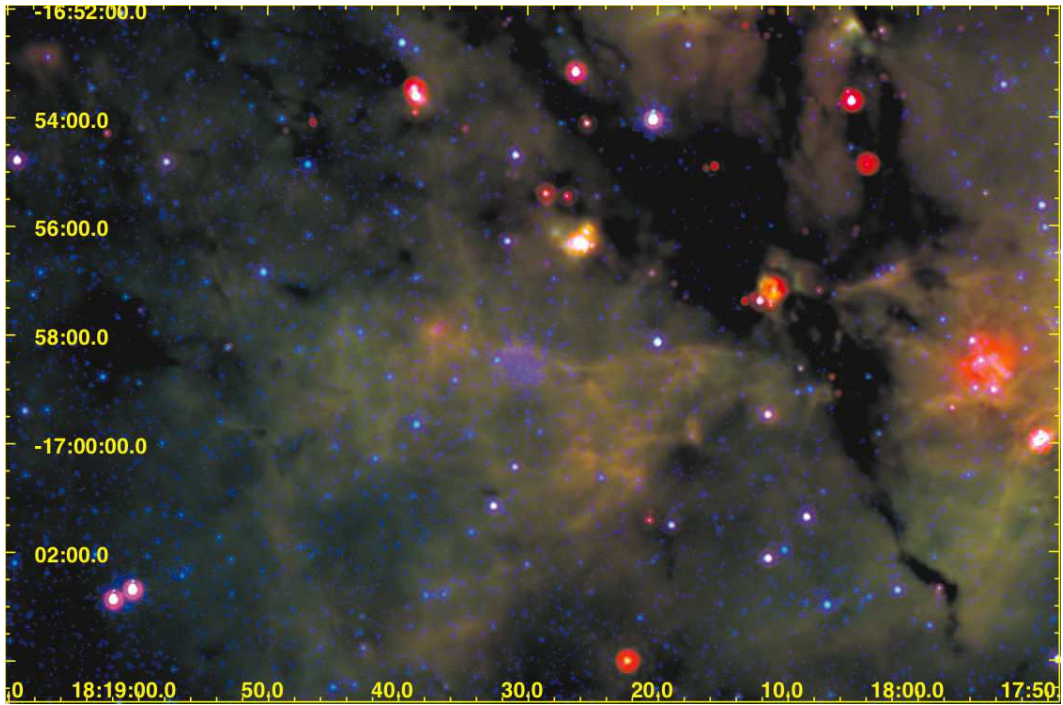}
\figcaption[newf2.eps]{\label{fig:fig3}
Three-color image using IRAC Bands 2 and 4 and MIPS 24 $\mu$m, showing the cluster
in the center of the image and its environment. Patchy extinction
is present throughout the image, and a large infrared dark cloud is present only
a few arcmin in projection from the cluster. This cloud may be associated with high
extinction toward the cluster.}
\end{figure}

\begin{figure} 
\plotone{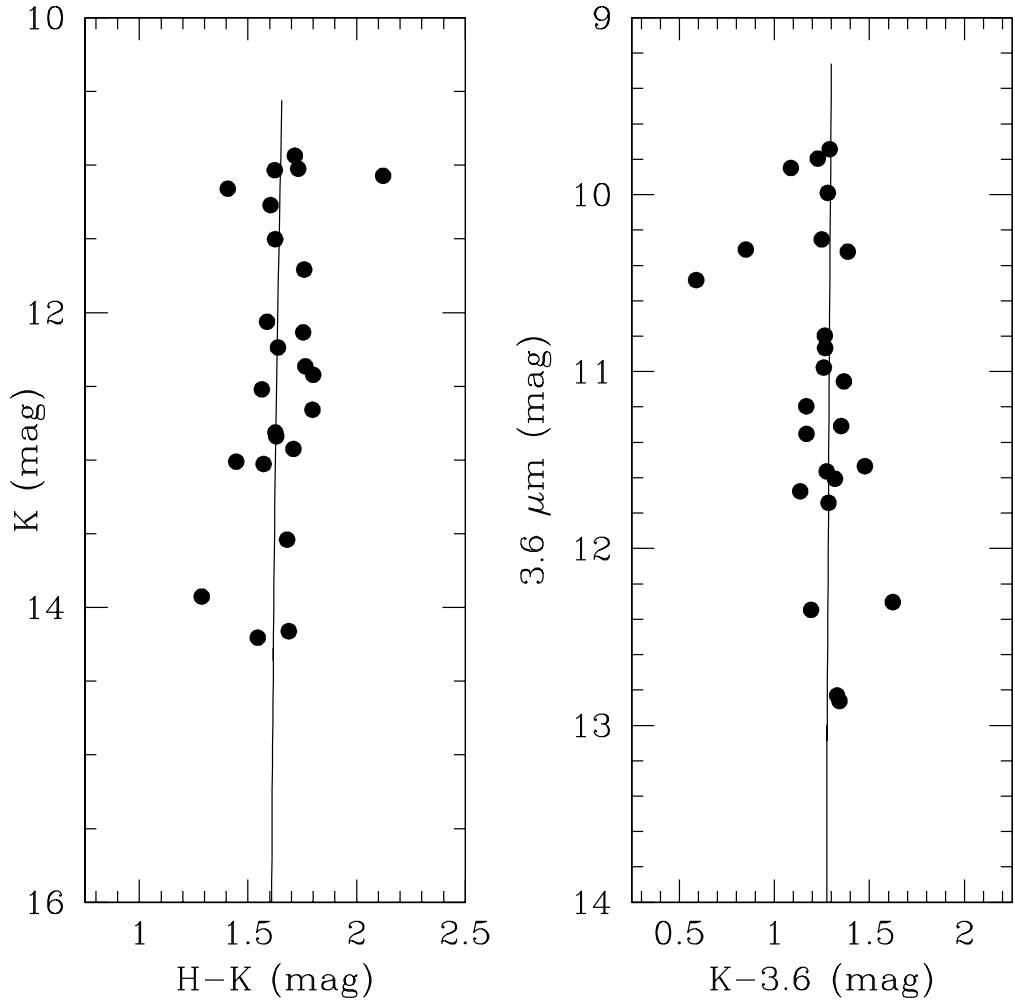} 
\figcaption[f4.eps]{\label{fig:fig4} 
$K$ vs.~$H-K$ and $3.6$ vs.~$K-3.6$ color-magnitude diagrams of Mercer 3. 12 Gyr, [$Z$/H] = $-2$ 
isochrones from Marigo \etal~(2008) assuming $E(B-V) = 7.7$ and a distance of 5 kpc are overplotted.} 
\end{figure}

\begin{figure}
\plotone{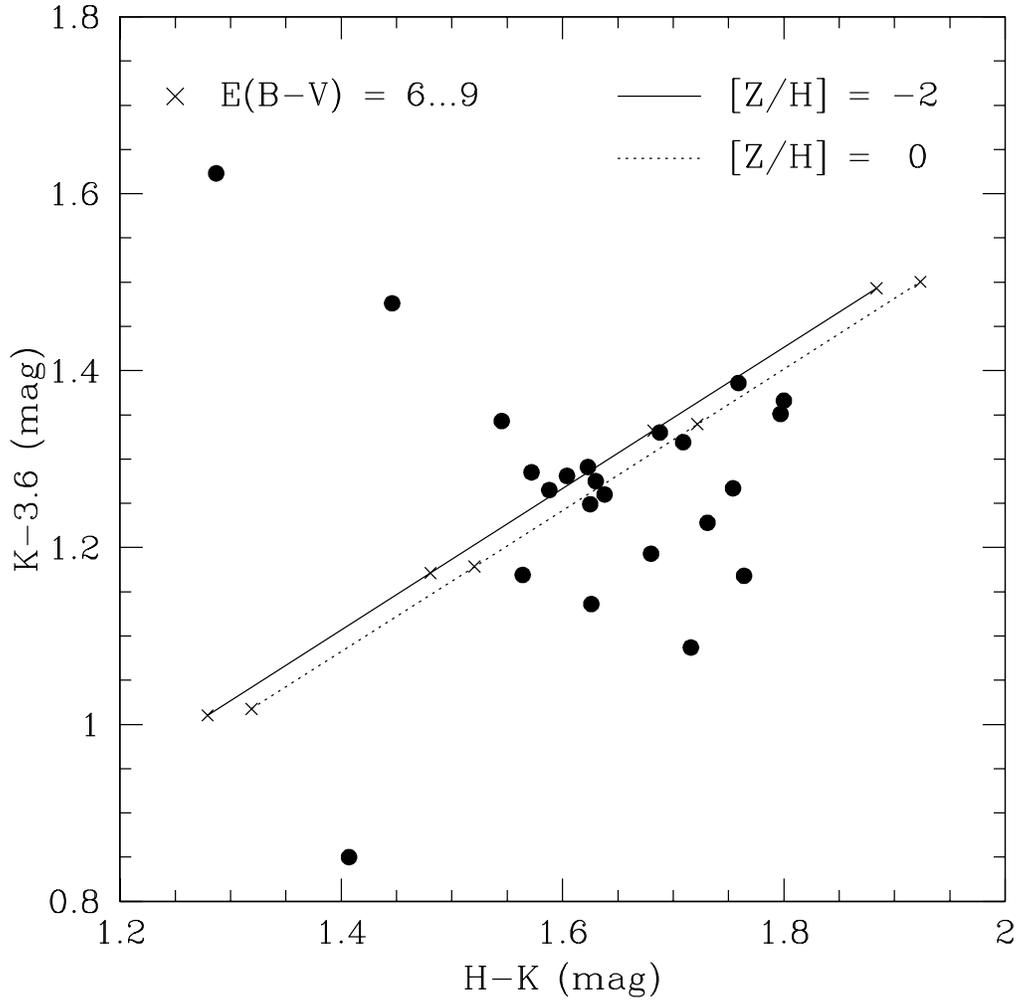}
\figcaption[f5.eps]{\label{fig:fig5}
$K-3.6$ vs.~$H-K$ color-color diagram of red giants in the cluster. Mean colors for the upper red giant 
branch using 12 Gyr isochrones with [$Z$/H] = $-2$ (solid line) and 0 (dotted line) are overplotted;
$E(B-V)$ ranges from 6 to 9 with crosses marking magnitude intervals. A value in the interval $7.5 
\la E(B-V) \la 7.8$ is favored.}
\end{figure}

\begin{figure} 
\plotone{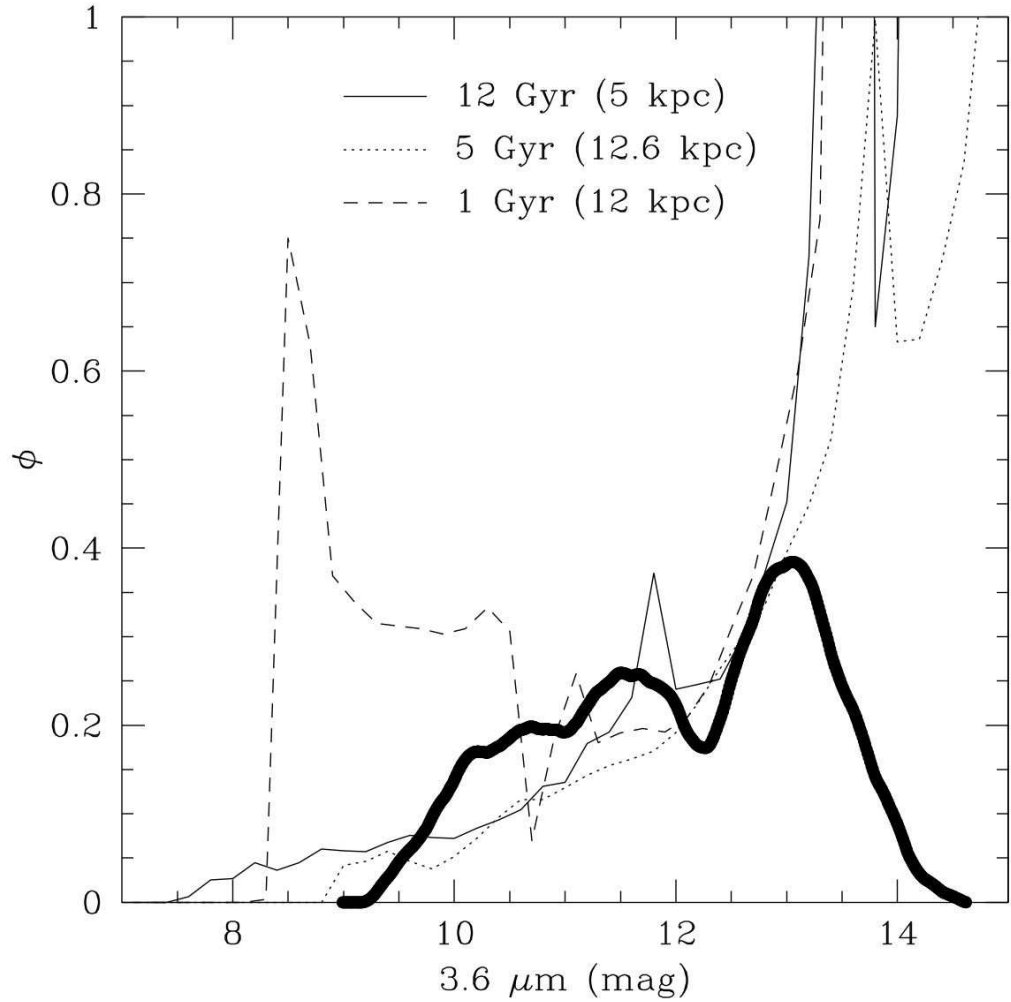} 
\figcaption[f6.eps]{\label{fig:fig6} Observed 3.6 $\mu$m luminosity function 
compared to theoretical luminosity functions of solar metallicity from Marigo \etal~(2008). These have ages of 12 
Gyr (solid), 5 Gyr (dotted), and 1 Gyr (short dashed). Ages younger than a few Gyr are disfavored 
because of the lack of red supergiants.}
\end{figure}


\begin{thebibliography}{}
\bibitem[Benjamin et al.(2003)]{2003PASP..115..953B} Benjamin, R.~A., et al.\ 2003, \pasp, 115, 953 
\bibitem[Bohlin et al.(1978)]{1978ApJ...224..132B} Bohlin, R.~C., Savage, B.~D., \& Drake, J.~F.\ 1978, \apj, 224, 132 
\bibitem[Bonatto \& Bica(2008)]{2008A&A...479..741B} Bonatto, C., \& Bica, E.\ 2008, \aap, 479, 741 
\bibitem[Bonatto et al.(2007)]{2007MNRAS.381L..45B} Bonatto, C., Bica, E., Ortolani, S., \& Barbuy, B.\ 2007, \mnras, 381, L45 
\bibitem[Clemens et al.(1986)]{1986ApJS...60..297C} Clemens, D.~P., Sanders, D.~B., Scoville, N.~Z., \& Solomon, P.~M.\ 1986, \apjs, 60, 297 
\bibitem[Harris(2001)]{2001stcl.conf..223H} Harris, W.~E.\ 2001, Saas-Fee Advanced Course 28: Star Clusters, 223 
\bibitem[Hurt et al.(2000)]{2000AJ....120.1876H} Hurt, R.~L., Jarrett, T.~H., Kirkpatrick, J.~D., Cutri, R.~M., Schneider, S.~E., Skrutskie, M., \& van Driel, W.\ 2000, \aj, 120, 1876 
\bibitem[Jackson et al.(2006)]{2006ApJS..163..145J} Jackson, J.~M., et al.\ 2006, \apjs, 163, 145 
\bibitem[Kobulnicky \& Skillman(2008)]{2008AJ....135..527K} Kobulnicky, H.~A., \& Skillman, E.~D.\ 2008, \aj, 135, 527 
\bibitem[Kobulnicky et al.(2005)]{2005AJ....129..239K} Kobulnicky, H.~A., et al.\ 2005, \aj, 129, 239 
\bibitem[Koposov et al.(2007)]{2007ApJ...669..337K} Koposov, S., et al.\ 2007, \apj, 669, 337 
\bibitem[Maraston(2005)]{2005MNRAS.362..799M} Maraston, C.\ 2005, \mnras, 362, 799 
\bibitem[Marigo et al.(2008)]{2008A&A...482..883M} Marigo, P., Girardi, L., Bressan, A., Groenewegen, M.~A.~T., Silva, L., \& Granato, G.~L.\ 2008, \aap, 482, 883 
\bibitem[Mercer et al.(2005)]{2005ApJ...635..560M} Mercer, E.~P., et al.\ 2005, \apj, 635, 560 
\bibitem[Worthey(1994)]{1994ApJS...95..107W} Worthey, G.\ 1994, \apjs, 95, 107 
\end{thebibliography}
\end{document}